\documentclass{INTERSPEECH2023}
\usepackage{hyperref}
\usepackage{caption}
\usepackage{subcaption}

\usepackage{microtype}
\usepackage{booktabs}
\usepackage{tabularx}
\usepackage{multirow}

\newcolumntype{C}{>{\centering\arraybackslash}X}
\newcolumntype{L}{>{\raggedright\arraybackslash}X}
\newcolumntype{R}{>{\raggedleft\arraybackslash}X}
\newcolumntype{P}[1]{>{\raggedright\arraybackslash}p{#1}}
\usepackage{siunitx}
\usepackage{etoolbox}                           \newcommand{\ubold}{\fontseries{b}\selectfont}  \robustify\ubold                                \newcommand{\tablecaptionsep}{\vspace*{0pt}}

\interspeechcameraready

\def\modelname{{kNN-VC}}

\title{Voice Conversion With Just Nearest Neighbors}

\renewcommand{\thefootnote}{\fnsymbol{footnote}}
\name{Matthew Baas$^{\scriptstyle\dagger}$\protect\footnote[2]{Equal contribution}, Benjamin van Niekerk$^{\scriptstyle\dagger}$\protect\footnote[2]{Equal contribution}, Herman Kamper}

\address{
  MediaLab, Department of Electrical \& Electronic Engineering, Stellenbosch University, South Africa
}
\email{20786379@sun.ac.za, benjamin.l.van.niekerk@gmail.com, kamperh@sun.ac.za}

\usepackage{hyperref}
\begin{document}

\maketitle
\renewcommand{\thefootnote}{\fnsymbol{footnote}}
\ifinterspeechfinal
\footnotetext[2]{Equal contribution.}
\else
\footnotetext[2]{[redacted for blind submission]}
\fi
\renewcommand*{\thefootnote}{\arabic{footnote}}
 
\begin{abstract}
Any-to-any voice conversion aims to transform source speech into a target voice with just a few examples of the target speaker as a reference.
Recent methods produce convincing conversions, but at the cost of increased complexity -- making results difficult to reproduce and build on.
Instead, we keep it simple.
We propose k-nearest neighbors voice conversion~\mbox{(kNN-VC)}: a straightforward yet effective method for any-to-any conversion.
First, we extract self-supervised representations of the source and reference speech. 
To convert to the target speaker, we replace each frame of the source representation with its nearest neighbor in the reference.
Finally, a pretrained vocoder synthesizes audio from the converted representation.
Objective and subjective evaluations show that \modelname{} improves speaker similarity with similar intelligibility scores to existing methods.
Code, samples, trained models: {\footnotesize \url{https://bshall.github.io/knn-vc}}.

\end{abstract}

\section{Introduction}

The goal of voice conversion is to transform source speech into a target voice, keeping the content unchanged~\cite{Mohammadi2017vc_def}.
We can categorize voice conversion systems by how restrictive the set of source and target speakers are.
The most general case is any-to-any conversion~\cite{vc_categories_liu2021any}, where the source and target speakers are unseen during training.
Here the goal is to convert to a target voice with just a few examples as a reference.
Recent any-to-any methods improve naturalness and speaker similarity, but at the cost of increased complexity~\cite{vcc2020, freevc}.
As a result, training and inference have become more costly, making improvements difficult to evaluate and build on.
We ask whether complexity is necessary for high-quality voice conversion.

To answer this question, we introduce k-nearest neighbors voice conversion (\modelname{}): a simple and robust method for any-to-any conversion.
Instead of training an explicit conversion model, we simply use k-nearest neighbors regression~\cite{knn_fix1985discriminatory}.
First, we use a self-supervised speech representation
model to extract feature sequences for both the source and reference utterances.
Next, we convert to the target speaker by replacing each frame of the source representation with its nearest neighbor in the reference.
Since certain self-supervised representations capture phonetic similarity~\cite{chen2022wavlm}, the idea is that the matched target frames would have the same content as the source.
Finally, we vocode the converted features with a neural vocoder to obtain the converted speech.

Focusing on any-to-any conversion, we compare \modelname{} to several state-of-the-art systems~\cite{freevc, vqmivc_wang21n_interspeech,yourtts_v162-casanova22a}.
Despite its simplicity, \modelname{} matches or improves intelligibility and speaker similarity in both subjective and objective evaluations. 
We also run ablations to understand how target data size and vocoder training decisions affect conversion quality.

Our main contribution is to show that complexity is not necessary for any-to-any voice conversion -- just applying nearest neighbors to self-supervised features gives convincing results.

\begin{figure*}[t!]
\centering
\centerline{\includegraphics[width=0.74\linewidth]{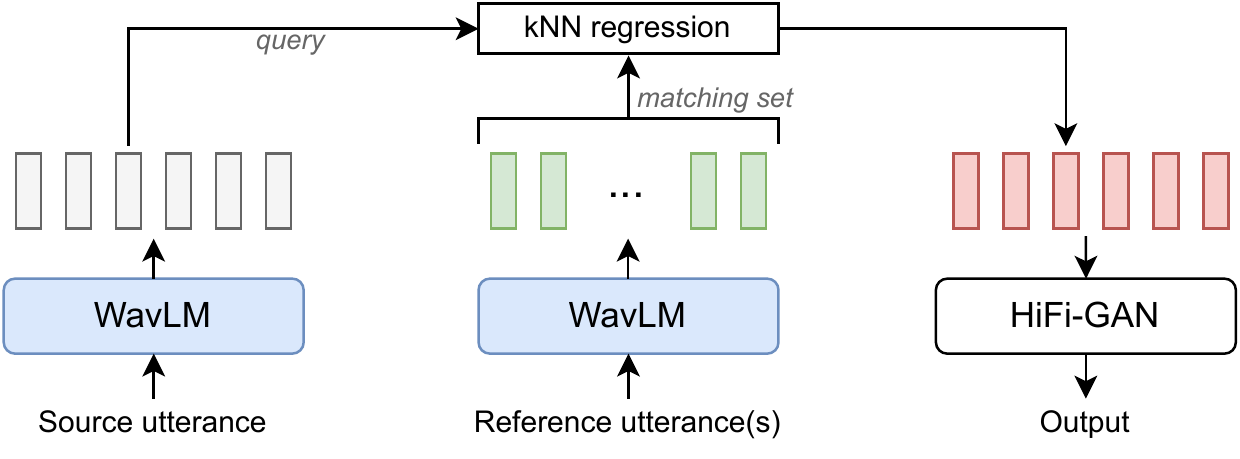}}
\caption{
    \modelname{}:
    The source and reference utterance(s) are encoded into self-supervised features from a pretrained WavLM model \cite{chen2022wavlm}.
    Each source feature is assigned to the mean of the $k$ closest features from the reference. 
    The resulting feature sequence is then vocoded using HiFi-GAN     to arrive at the converted waveform output.
}
\label{fig:1_model_arch}
\end{figure*}

\section{Related work}

Typical voice conversion systems try to separate speaker identity from content.
The speaker information can then be replaced
to convert to a target voice.
However, learning disentangled representations is challenging~\cite{vcc2020}.
Recent systems require complex information bottlenecks~\cite{freevc, vqmivc_wang21n_interspeech, autovc}, normalization methods~\cite{instancenormvc}, or data augmentation \cite{freevc}, but still struggle with speaker similarity.
Alternatively, some methods use text annotations as an intermediate information bottleneck to improve disentanglement~\cite{vc_categories_liu2021any,yourtts_v162-casanova22a}.
However, transcribing large speech corpora is costly and time-consuming.
Additionally, the bottleneck imposed by text can remove prosody information which is beneficial in voice conversion applications.
We therefore
focus on text-free methods in this paper.

Concatenative voice conversion is another line of work related to our proposed approach~\cite{concatenative_vc_2007, cuta_concatenative_vc}.
The idea in concatenative conversion is to stitch together units of target speech that match the source content.
By constructing the output exclusively from target speech, concatenative methods ensure good speaker similarity.
A key component in concatenative systems is unit selection, which generally requires parallel data~\cite{cuta_concatenative_vc} or time-aligned transcriptions~\cite{concatenative_text}.
Our method can also be seen as a unit selection approach. But instead of using human-defined units, we use features from a self-supervised speech model.
Recent studies have shown that self-supervised representations linearly predict many properties of speech~\cite{wav2vec_weight_analysis_2021,wavlm_prosody_weights2023}.
Specifically, similar features indicate shared phonetic content~\cite{zerospeech}.
This leads to our research question --
are complex methods still necessary for high-quality voice conversion given 
\mbox{these improved speech representations?}

\section{\modelname{}}

To answer this question, we propose k-nearest neighbors voice conversion (kNN-VC).
Figure~\ref{fig:1_model_arch} shows an overview of the method, which follows an encoder-converter-vocoder structure~\cite{vc_overview_2021}. 
First, the encoder extracts self-supervised representations of the source and reference speech.
Next, the converter maps each source frame to its nearest neighbor in the reference.
Finally, the vocoder generates an audio waveform from the converted features.
Each of these components is described below.

\subsection{Encoder}

\modelname{} starts by extracting the feature sequence of the source utterance, which we call the \textit{query sequence} (Figure~\ref{fig:1_model_arch}).
We also extract the feature sequences of one or more utterances from the target speaker, which get shuffled together into a large pool of self-supervised feature vectors.
We call this bag-of-vectors the \textit{matching set}.
The goal of this encoder is to extract representations where nearby features have similar phonetic content.
Recent self-supervised models \cite{wav2vec2.0, chen2022wavlm} are good candidates since they score well on phone discrimination tests, i.e., they encode instances of the same phone closer to each other than to different phones~\cite{zerospeech}.
The encoder model is never fine-tuned or trained further for \modelname{} -- it is only used to extract features.

\subsection{k-nearest neighbors matching}

To convert to the target speaker, we apply k-nearest neighbors~(kNN) regression to every vector in the query sequence.
Specifically, we replace each query frame with the average of its k-nearest neighbors in the matching set.
Since similar self-supervised speech features share phonetic information~\cite{chen2022wavlm,wavlm_prosody_weights2023}, performing kNN preserves the content of the source speech while converting speaker identity.
Similarly to concatenative methods, we construct the converted query directly from target features, guaranteeing good speaker similarity.
The kNN regression algorithm~\cite{knn_fix1985discriminatory} is also non-parametric and requires no training, making this method easy to implement.

\subsection{Vocoder}

The vocoder translates the converted features into an audio waveform.
Instead of conditioning on spectrograms, we adapt a conventional vocoder to take self-supervised features as input.
However, there is a mismatch between inputs during training and inference.
For inference, we condition the vocoder on kNN outputs, i.e., the mean of features selected from the matching set.
We pick these features from various points in time with different phonetic contexts, leading to inconsistencies between adjacent frames.
Section~\ref{subsec:5.2_ablation} shows that these inconsistencies cause artifacts, reducing the intelligibility of converted speech.

\subsection{Prematched vocoder training}
\label{sec:3_prematching}

To address this issue, we propose \textit{prematched training}.
Specifically, we reconstruct the vocoder training set using k-nearest neighbors.
Using each training utterance as a query, we build a matching set using the remaining utterances from the same speaker.
Then we apply kNN regression to reconstruct the query sequence using the matching set.
We train the vocoder to predict the original waveforms from these prematched features.
The idea is to improve robustness by training the vocoder on data resembling what it encounters during inference.
In Section~\ref{subsec:5.2_ablation}, we show that prematched vocoder training improves intelligibility and speaker similarity.

\section{Experimental setup}

To determine whether our new non-parametric voice conversion method can compete with recent neural network 
conversion models, we run two experiments on the LibriSpeech development and test sets.
The first experiment compares kNN-VC to state-of-the-art voice conversion systems.
The second investigates the effect of target data size and prematched vocoder training.
For both experiments, the source and target speakers are unseen during training.
The LibriSpeech development and test clean sets consist of 40 speakers apiece, each contributing about 8 minutes of 16~kHz spoken English~\cite{panayotov2015librispeech}.
The diversity of speakers and audio quality in LibriSpeech make this a challenging benchmark for any-to-any voice conversion.

\subsection{\modelname{} implementation}

\textbf{Encoder}: We use the pretrained WavLM-Large encoder from~\cite{chen2022wavlm} to extract features for the source and reference utterances.
In preliminary experiments, we used features from later layers (22, 24, and the mean of the last several layers), which perform well on linear phone recognition tasks~\cite{chen2022wavlm}.
The idea was to improve nearest neighbors mapping by including more content information.
However, this led to worse pitch and energy reconstruction.
Recent work~\cite{wavlm_prosody_weights2023} confirms that later layers
give poorer predictions of pitch, prosody, and speaker identity. Based on these observations, we found that using a layer with high correlation with speaker identification -- layer 6 in WavLM-Large -- was necessary for good speaker similarity and retention of the prosody information from the source utterance.
So, for all the following experiments, we use features extracted from layer 6 of WavLM-Large, which produces a single vector for every 20~ms of 16~kHz audio.\\

\noindent \textbf{kNN regression}: 
For all our experiments, we set $k=4$ with a uniform weighting and use cosine distance to compare features.
In preliminary experiments, we found that \modelname{} is quite robust to a range of values around $k=4$. 
Namely, when more reference audio is available (e.g. $\geq$10~mins), the conversion quality may even be improved by using larger values of $k$ (in the order of $k=20$). \\

\noindent \textbf{Vocoder}: We train the HiFi-GAN V1 architecture from \cite{hifi-gan}.
We modify it
to accept the 1024-dimensional input vectors from WavLM and to vocode 16~kHz audio using 128-dimensional mel-spectrograms
with a 10~ms hop length and 64~ms Hann window.
We train the model using the same optimizer, steps and other hyperparameters as \cite{hifi-gan} on the LibriSpeech train-clean-100 dataset.
We train two
variants: one trained on pure WavLM-Large layer 6 features, and one trained on prematched layer 6 features (as explained in Section~\ref{sec:3_prematching}).
With these design choices, inference with 8 minutes of reference audio on a consumer 8GB VRAM GPU is faster than real-time.

\subsection{Baselines}

We compare \modelname{} to three any-to-any voice conversion systems: 
VQMIVC\footnote{\scriptsize \url{https://github.com/Wendison/VQMIVC}}, 
FreeVC\footnote{\scriptsize \url{https://github.com/OlaWod/FreeVC}}, and YourTTS\footnote{\scriptsize \url{https://github.com/Edresson/YourTTS}} (in text-free voice-conversion mode).
VQMIVC uses vector quantization with mutual information 
minimization
to disentangle speaker identity from content~\cite{vqmivc_wang21n_interspeech}.
FreeVC combines a variational autoencoder with data augmentation to discard speaker information~\cite{freevc}.
YourTTS uses text transcriptions to create an intermediate information bottleneck during training~\cite{yourtts_v162-casanova22a}.
To convert to the target speaker, all three models are conditioned on speaker embeddings extracted from the reference speech using a speaker encoder.
During inference, we average the speaker embeddings across all utterances from the target speaker to maximize performance.
We use the publicly available, pretrained models provided by the respective authors as the baselines.

\subsection{Evaluation metrics}

We assess naturalness, intelligibility and speaker similarity in subjective and objective evaluations.
We report results on the LibriSpeech test-clean subset which consists of unseen speakers not contained in any other subset of LibriSpeech.\\

\noindent\textbf{Objective evaluations}:
To construct an evaluation set, we sample 200 utterances (5 per speaker) from the LibriSpeech test-clean set. 
We convert each utterance to the 39 other speakers, giving a total of 7800 outputs per model.
Following previous studies~ \cite{vcc2020,vqmivc_wang21n_interspeech,vc_categories_liu2021any}, we evaluate  intelligibility by calculating the word/character error rate (W/CER) of
an ASR system applied to the converted speech.
Lower error rates indicate better intelligibility.
We use the trained Whisper-base ASR model~\cite{whisper_radford2022robust} with default decoding parameters \mbox{to transcribe the converted utterances.}

To measure speaker similarity, we follow~\cite{softvc} and calculate an equal error rate (EER) using a trained speaker verification system \cite{xvector}.
Given an input utterance, the verification system outputs a x-vector representing the speaker identity.
We compute the cosine similarity between x-vectors to determine whether two utterances are from the same speaker.
First, we record similarity scores for each converted sample paired with an enrollment utterance from the target speaker.
Then, we obtain an equal number of `ground-truth' similarity scores by sampling target speaker utterances and computing the similarity to a different enrollment target speaker utterance. Finally, we calculate an EER over the combined set of scores, assigning a label of 1 to the ground-truth pairs and 0 to pairs containing converted speech.
A higher EER indicates that it is more difficult to distinguish converted speech from genuine examples of the target speaker, i.e., higher EER corresponds to better speaker similarity, with a maximum of 50\%.\\

\noindent \textbf{Subjective evaluations}:
Using Amazon Mechanical Turk, we run subjective evaluations to measure naturalness and
speaker similarity.
For naturalness, we report mean opinion scores (MOS) on a 1-to-5 scale.
We sample and evaluate 60 utterances for each model, adding an equal number of ground-truth examples from the LibriSpeech test-clean set.
We filter out ratings from listeners who judge the ground-truth examples as very unnatural (a MOS of 1), resulting in a total of 2144 ratings from 61 listeners.

For speaker similarity, we follow the protocol from the Voice Conversion Challenge 2020 \cite{vcc2020}:
given a pair of utterances, we ask listeners to rate the similarity between the speakers on a 1-to-4 scale.
For the evaluation, we sample 20 source utterance from the LibriSpeech test-clean set.
We convert each utterance to three target speakers using each model, resulting in 60 examples per model.
We ask raters to evaluate the similarity between the converted outputs and a genuine example of the target speaker.
Again, we again filter out raters that judge the ground-truth target speaker pair as very different (a SIM of 1),
resulting in a total of 1636 ratings over 208 listeners. 
Based on these ratings, we report a mean similarity (SIM) score for each model.

\section{Results} \label{sec:5_results}

We report the results of the comparative experiment in Section~\ref{sec:5.1_comparitive}, followed by the ablation experiment in Section~\ref{subsec:5.2_ablation}.
In the first experiment, we test the voice conversion systems using the maximum available target data (about 8 minutes of audio per speaker).
For \modelname{}, we use all the target data as our matching set.
For the baselines, we average the speaker embeddings of each target utterance.
The second experiment evaluates kNN-VC as we vary the target data size.
We also investigate how prematched training affects intelligibility and speaker similarity.

\setlength{\tabcolsep}{3.6pt}
\begin{table}[!b]
    \renewcommand{\arraystretch}{1.2}
        \centering
    \caption{
        Results measuring the intelligibility (W/CER), naturalness (MOS) and speaker similarity (EER, SIM) on the LibriSpeech test-clean subset. Subjective MOS and SIM values with 95\% confidence intervals are shown.
    }
    \tablecaptionsep
    \eightpt
    \label{tab:1_headline_results}
    \begin{tabularx}{1.0\linewidth}{@{}
        L
        S[table-format=2.2]
        S[table-format=2.2]
        S[table-format=2.2]
        cc
        @{}}
    \toprule
    Model & {WER$\ \downarrow$} & {CER $\ \downarrow$} & {EER $\ \uparrow$} & {MOS $\ \uparrow$} & {SIM$\ \uparrow$} \\
    \midrule
    \textit{Testset topline} & 5.96 & 2.38 & {-} & 4.24$\pm$0.07 & 3.19$\pm$0.09 \\
        \addlinespace
    VQMIVC \cite{vqmivc_wang21n_interspeech} & 59.46 & 37.55 & 2.22 & 2.70$\pm$0.11 & 2.09$\pm$0.12 \\
    YourTTS \cite{yourtts_v162-casanova22a} & 11.93 & 5.51 & 25.32 & 3.53$\pm$0.09 & 2.57$\pm$0.12 \\
    FreeVC \cite{freevc} & 7.61 & 3.17 & 8.97 & \ubold 4.07$\pm$0.07 & 2.38$\pm$0.11 \\
    \modelname{} & \ubold 7.36 & \ubold 2.96 & \ubold 37.15 & \ubold 4.03$\pm$0.08 & \ubold 2.91$\pm$0.11 \\
    \bottomrule
    \end{tabularx}
\end{table}

\subsection{Voice conversion}
\label{sec:5.1_comparitive}

Table~\ref{tab:1_headline_results} reports intelligibility, naturalness, and speaker similarity results for each model.
For \modelname{}, we use the HiFi-GAN trained on prematched data. We observe that \modelname{} achieves similar naturalness (MOS) and intelligibility (W/CER) to the best baseline, FreeVC.
However, \modelname{} significantly improves speaker similarity (EER and SIM). Given the simplicity of \modelname{}, these results confirm our research hypothesis: increased complexity is not necessary for high-quality voice conversion.

A stronger conclusion is that
our simple kNN method, combined with self-supervised speech features, leads to improved speaker similarity compared to existing models.
A major benefit of kNN-VC is that it does not require an explicit speaker embedding model, in contrast to all the \mbox{baselines we compare to here.}

Because we do not rely on a speaker embedding model, an interesting question is: how far away can the reference utterances be from the training distribution?
We leave a full investigation to future work, but include conversions to unseen languages, whispered speech, and even non-speech sounds on our demo page:
{\footnotesize \url{https://bshall.github.io/knn-vc}}.
For example, in our cross-lingual demo, we convert German source speech to a Japanese target speaker.
The conversion is intelligible despite the system having only seen English during its design and training.

\begin{figure}[t!]
\centering
\centerline{\includegraphics[width=0.98\linewidth]{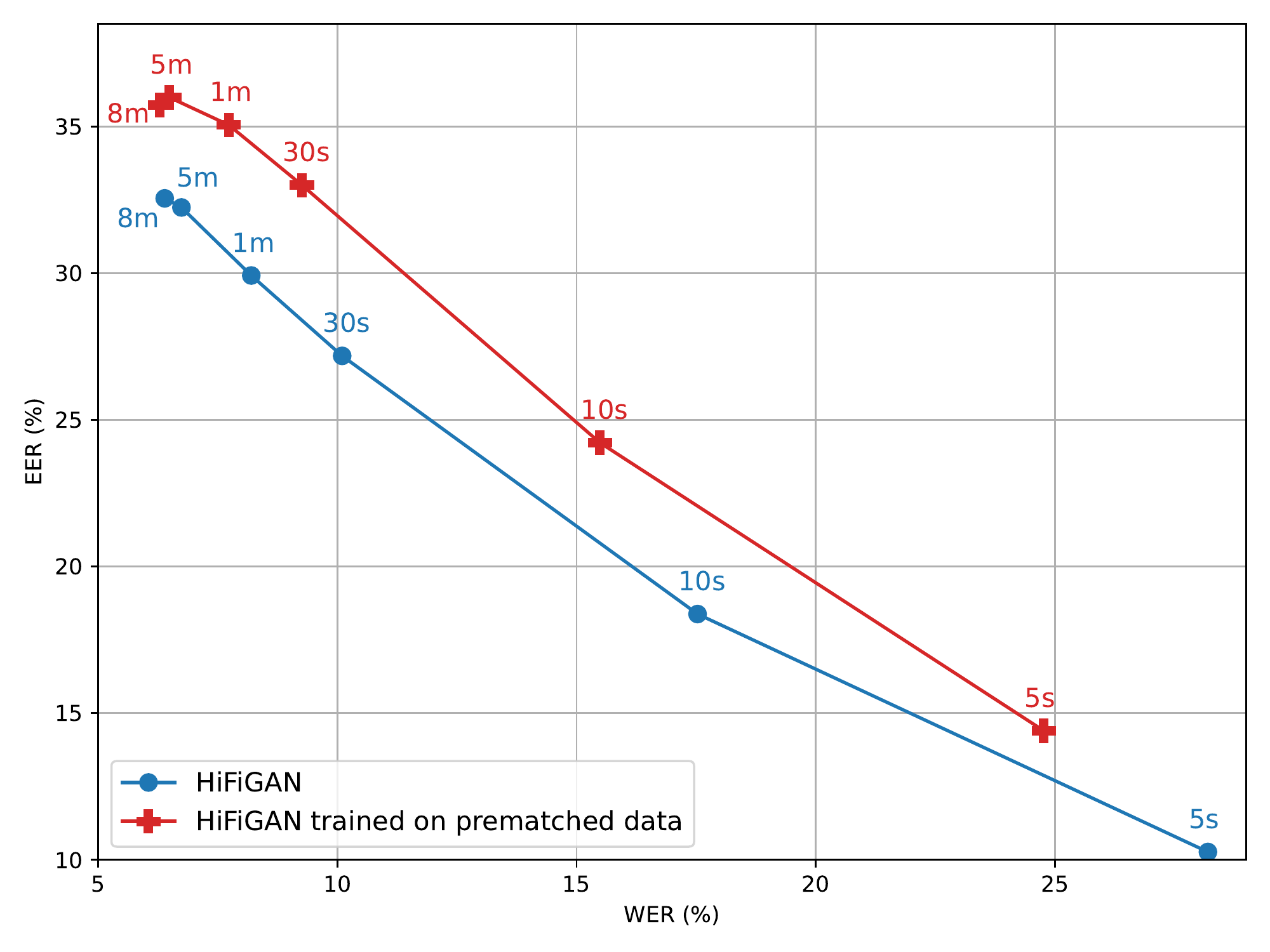}}
\caption{
    Intelligibility (WER, lower is better) vs speaker similarity (EER, higher is better) for varying amounts of target speaker data on the LibriSpeech dev-clean subset.
        The optimal result is to be in the upper-left corner.
    }
\label{fig:2_ablation}
\end{figure}

\subsection{Ablation: Prematching and amount of reference data}\label{subsec:5.2_ablation}

While the main research question is answered,
we would like to know
how much of the improvements come from HiFi-GAN being trained on prematched data (Section~\ref{sec:3_prematching}), and how much the quantity of target speaker data influences intelligibility and speaker similarity. 
To elucidate this, Figure 2 plots WER against EER across different target speaker data sizes for both HiFi-GAN variants. \\

\noindent \textbf{Prematched data}: 
To recap from Section~\ref{sec:3_prematching}, prematching works by mapping each HiFi-GAN training utterance's WavLM features to the nearest $k$ vectors from other utterances of the same speaker (rather than just training on WavLM features directly). The higher EER and lower WER in Figure~\ref{fig:2_ablation} shows that prematching gives
a non-negligible improvement, regardless of the amount of target speaker data. \\ 
\noindent \textbf{Amount of reference data}: 
It is clear from Figure~\ref{fig:2_ablation} that as
amount of target speaker speech
decreases,
both the intelligibility and speaker similarity of the converted utterance deteriorates.
This is to be expected since in the extreme case of e.g.\ 5 or 10~seconds of reference audio, it is unlikely that the matching set 
would cover
all phone or biphones necessary for smooth conversion.
We see, however, that the performance gains diminishes after a certain point:
using the maximum speaker data in LibriSpeech (roughly 8 minutes per speaker)
gives
very similar performance to only using 5 minutes of target speaker data.
Again, this is likely because -- beyond a certain point -- there are enough vectors in the matching set to cover all possible phones and biphones present in the source utterance. \\

\noindent \textbf{Comparison to baselines}: Even with the plain HiFi-GAN vocoder not trained on prematched data, the scores that \modelname{} achieves in Figure~\ref{fig:2_ablation} are still competitive with existing models in terms of intelligibility and still superior in terms of speaker similarity (comparing to the scores in Table~\ref{tab:1_headline_results}).
This demonstrates that, while training the vocoder on prematched data helps, using the simplest setup of a pretrained self-supervised speech model and vocoder together with kNN regression is still a powerful any-to-any voice conversion approach. However, with limited target data (less than 30s),  intelligibility and target speaker similarity decrease to a point where the more complex baselines perform better (Table~\ref{tab:1_headline_results}).

\section{Conclusions}

In this work, we investigated whether complex methods are necessary 
for any-to-any
voice conversion. Using representations from a self-supervised speech model, we 
proposed
\modelname{}: a conversion approach based on the simple kNN algorithm. 
We compared \modelname{} to several recent high-performing voice conversion models in a challenging any-to-any conversion scenario, 
and found
that we achieve similar intelligibility and naturalness scores while improving speaker similarity.
This confirms our main hypothesis that voice conversion can be done with simpler methods that make use of general self-supervised speech representations. We also introduced the idea of training a vocoder on prematched feature data to better approximate inference conditions, which improved performance.
To better understand the limits of our approach, we 
looked at the effect of the amount of
reference data on final voice conversion performance.
We found that, with as little as five seconds of target speaker audio, we can still retain moderate intelligibility and speaker similarity.

Because the core of our approach is the vanilla kNN algorithm and our code and methods are open-source,
we hope that \modelname{} can serve as a robust, easy-to-implement baseline for voice conversion.
Additionally, because \modelname{} is non-parametric, we include qualitative examples of its application to cross-lingual, whispered, and even human-to-animal voice conversion -- areas we hope to explore more thoroughly in future work.

\section{Acknowledgements}

\ifinterspeechfinal
All experiments were performed on Stellenbosch University’s High Performance Computing (HPC) GPU cluster. This work is supported in part by the National Research Foundation of South Africa (grant no. 120409).
M. Baas is funded by the Harry Crossley Foundation.
\else
All experiments were performed on [redacted for blind submission]. This work is supported in part by the [redacted for blind submission].
\fi

\newpage

\bibliographystyle{IEEEtran}
\bibliography{mybib}

\end{document}